\documentclass{article}
\usepackage{spconf,amsmath,graphicx}

\usepackage{enumitem}
\setlist{nosep, leftmargin=14pt}

\title{Evaluation of Augmentation methods in classifying Autism Spectrum Disorders from fMRI data with 3D Convolutional Neural Networks}
\name{Johan J\"onemo$^{\: a,c}$, David Abramian$^{\: a,c}$, Anders Eklund$^{\: a,b,c}$}
\address{$^a$Division of Medical Informatics, Department of Biomedical Engineering\\ $^b$Division of Statistics \& Machine Learning, Department of Computer and Information Science\\ $^c$Center for Medical Image Science and Visualization (CMIV) \\Link\"{o}ping University, Sweden }

\begin{document}
%
\maketitle
\begin{abstract}
Classifying subjects as healthy or diseased using neuroimaging data has gained a lot of attention during the last 10 years. Here we apply deep learning to derivatives from resting state fMRI data, and investigate how different 3D augmentation techniques affect the test accuracy. Specifically, we use resting state derivatives from 1,112 subjects in ABIDE preprocessed to train a 3D convolutional neural network (CNN) to perform the classification. Our results show that augmentation only provide minor improvements to the test accuracy.
\end{abstract}
\section{Introduction}
\label{sec:intro}

Ever since the emergence of magnetic resonance imaging (MRI) in the 1980's, the absence of ionizing radiation
and the flexibility of the acquisition procedure have made this an increasingly
important imaging modality in the clinical sciences. The lack of contrast between
different tissues in the brain and the interference of the mineralized tissue
around it when using x-ray techniques, make MRI especially useful in neuroimaging.

While a wide variety of neurological conditions can be diagnosed with MRI, psychiatric
anomalies, however, have proven illusive to detect. Presumably this is because these
affect many systems distributed throughout the brain and their manifestations are
likely subtle as well as time variant. Functional MRI (fMRI) is a technique that seems
particularly suited to capture this information and several large collaborative efforts
have been made to collect and share (resting state) fMRI data, for different psychiatric conditions~\cite{poldrack2014}.

ABIDE~\cite{dimartino2014} is one such effort that make available data for 539 subjects
with Autism Spectrum Disorders (ASD) as well as 573 typical controls. We are using machine learning
in an endeavour to classify fMRI data according to the presence or absence of ASD. The
problem seems hard in that accuracies seldom rise to more than 70\% when the model
classifies unseen data~\cite{arbabshirani2017,thomas2020,heinsfeld2018}. While 1,112 subjects is a very large fMRI dataset, it is still small from a deep learning perspective (for example, the popular ImageNet database contains several million images). To further increase the size of the training dataset, and to make CNNs robust to transformations such as rotation, data augmentation is often used~\cite{shorten2019}. In our recent work we investigated what the best 3D augmentation is for brain tumor segmentation~\cite{cirillo2021}. In this paper we instead want to see if 3D augmentation can help train a better classifier as well as what kind of augmentation techniques work the best. Thomas et al.~\cite{thomas2020} recently used deep learning for the same ABIDE dataset, but do not mention anything about augmentation.

We have therefore implemented a 3D convolutional neural network (CNN) of modest size and trained it with resting state fMRI deriviatives from ABIDE, using no augmentation as well as using different augmentation strategies. Briefly, our results show some improvements with carefully
chosen augmentation strategies but the effects are in the order of magnitude of a percentage point. Out of the commonly used augmentation techniques that we evaluated, many proved to be deleterious in this context.

\section{Methods}
\label{sec:format}

\subsection{Data}

ABIDE preprocessed~\cite{craddock2013}\footnote{http://preprocessed-connectomes-project.org/abide/} shares preprocessed data from resting state fMRI in various forms. As all the preprocessing has been done, we can focus on the machine learning part, and other researchers can use the same preprocessed data to reproduce our findings. We downloaded 3D data of derivatives where the time dimension had been collapsed into different forms of statistics.

Specifically, we downloaded fMRI data (derivatives) processed with the Connectome Computation System (CCS) pipeline~\cite{xu2015} which performs slice timing correction,
motion realignment and global intensity normalisation. The data was cleaned from
confounders by regressing with the head movement parameters, the time dependent mean intensity
as well as regressors for linear and quadratic drift. Each time series was also band pass filtered (0.01 - 0.1 Hz). This preprocessing corresponds
to the strategy called global\_filt.
The data was furthermore registered to the MNI152 brain template using boundary based rigid body registration~\cite{greve2009} for functional to anatomical registration, and FLIRT and FNIRT for anatomical
to template registration~\cite{jenkinson2012}. 

After preliminary testing of the 10 calculated derivatives available in ABIDE preprocessed, we chose to use Regional Homogeneity (ReHo) derivative for comparing different augmentation strategies. Each derivative volume from the resting state fMRI data has a size of 61 x 73 x 61 voxels, which is fed into the 3D CNN described below. The 539 subjects with ASD and the 573 controls were split 70/15/15 into training, validation and test sets.

\vspace{-0.1cm}
\subsection{Deep learning}

The 3D CNN was implemented using Keras and consists of 3 convolutional layers (ReLU activation), max-pooling layers, a dense layer with 16 nodes, and a final one-node layer with sigmoid activation. The first and second convolutional layers contain 8 filters each (size 3 x 3 x 3), and the last convolutional layer uses 16 filters. The total number of trainable parameters in the 3D CNN is approximately $450$k. The CNN was trained with the Adam optimizer with a learning rate of $10^{-5}$. To prevent overfitting, early stopping was used with a patience of 50 epochs. The training was run until validation accuracy did not improve, and the model then restored to the state when the last improvement was seen. As an alternative, the models were also trained for 150 epochs with no conditional stopping. To obtain more robust estimates of the test accuracy, 10-fold cross validation was used and the mean test accuracy was calculated.

The average training time for a single fold were between five minutes and 2.5 hours - depending on the type of on-the-fly augmentation employed - with one Nvidia Tesla V100 graphics card for the early stopping models. For the training with a fixed number of epochs, the average single fold training time was at least 10 minutes but otherwise in the previously mentioned span. In the longer training runs it is unlikely that the computation speed was bounded by the speed of the graphics card, as the on-the-fly augmentations were done on the CPU. In total we trained 300 3D CNNs in order to compare all settings.
 
\vspace{-0.2cm}
\subsection{Augmentation}

There are many types of augmentation that can be useful in 3D. Rotation, flipping and scaling are common for training 2D CNNs, and can easily also be applied in 3D. Elastic deformations are common when training segmentation networks, but perhaps not as common for classification. Brightness augmentation can for example help if the data have been collected at several different MR scanners, as they normally generate data with different brightness~\cite{cirillo2021}. While 2D augmentation is included in many deep learning frameworks, the support for 3D augmentation is normally lacking, which forced us to make our own implementation. The 3D augmentation techniques tested in this study are: 
\begin{itemize}
    \item \textit{Flipping}: flipping of the x-axis or not, which seems to work better compared to flipping all axes. 
    \item \textit{Rotation}: rotation applied to each axis with angles randomly chosen from a uniform distribution with range between -7.5 and 7.5 degrees, -15 and 15 degrees, -30 and 30 degrees or -45 and 45 degrees.
    \item \textit{Scale}: scaling applied to each axis by a factor randomly chosen from a uniform distribution with range \(\pm10\)\% or \(\pm20\)\%. 
    \item \textit{Brightness}: power-law \(\gamma\) intensity transformation with its parameters gain (\(g\)) and \(\gamma\) chosen randomly between 0.8 - 1.2 from a uniform distribution. The intensity (\(I\)) is randomly changed according to the formula: \(I_{new} = g\cdot I^{\gamma}\).
    \item \textit{Elastic deformation}: elastic deformation with square deformation grid with displacements sampled from from a normal distribution with standard deviation \(\sigma = 2\), 4, 6, or 8 voxels~\cite{ronneberger2015}, where the smoothing is done by a spline filter with order 3 in each dimension. 
\end{itemize}

To investigate the effect of combining different types of augmentation, we also trained the CNN with the two best performing augmentation approaches.

\vspace{-0.25cm}
\section{Results}
\label{sec:results}

\begin{figure*}[tbp]
    \centering
    \includegraphics[width=0.88\textwidth]{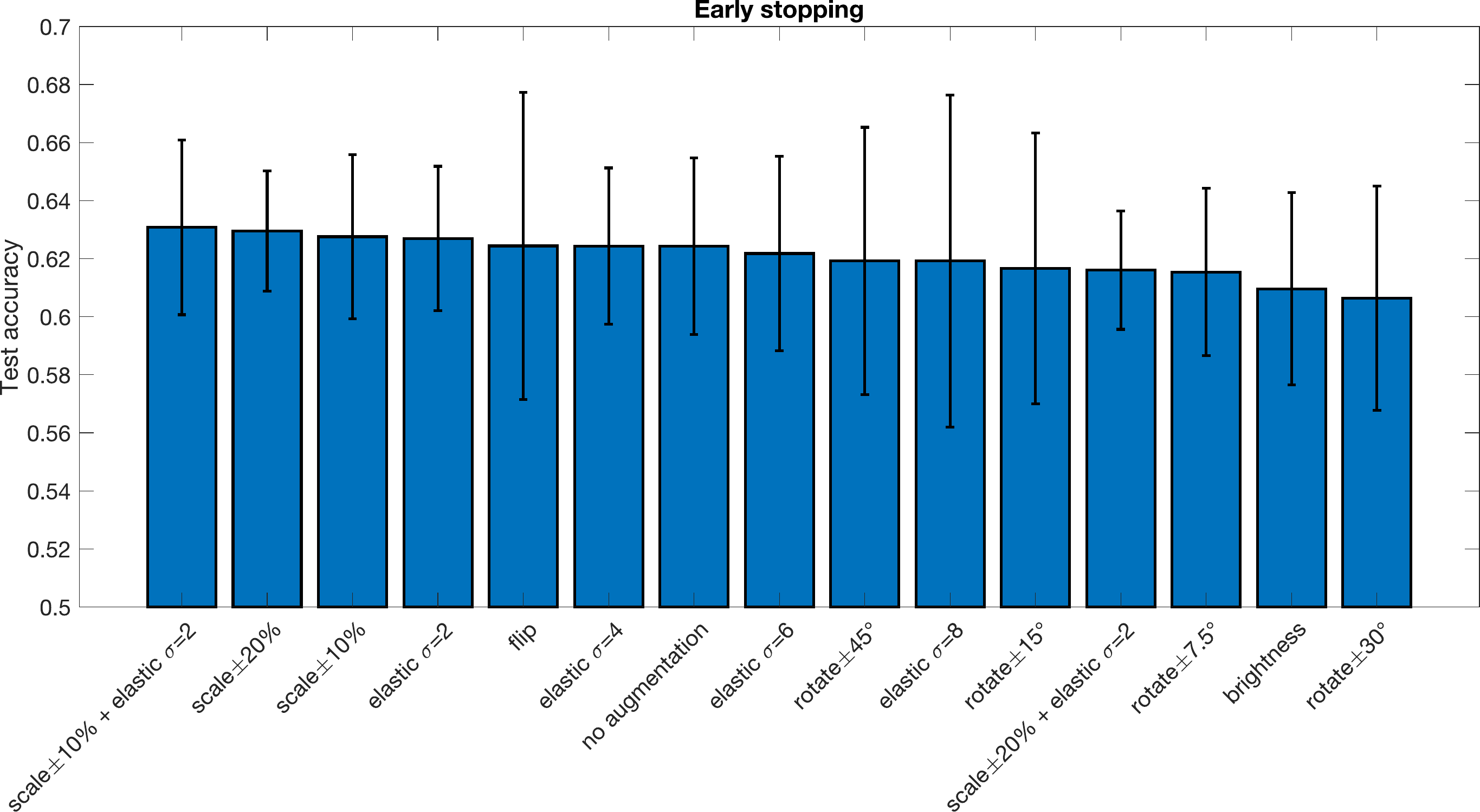}
    \caption{Test accuracy for classifying subjects as healthy or diseased for the ABIDE dataset, for different data augmentation approaches. The error bar represents the standard deviation over the 10 cross-validation folds. Note that half of the augmentation approaches result in a test accuracy that is lower compared to the baseline model trained without augmentation, but overall the differences are small. These results were obtained when using early stopping. Compared to no augmentation, the best augmentation approach increases the test accuracy by 0.6 percentage units.}
    \label{fig:results1}
\end{figure*}

\begin{figure*}[tbp]
    \centering
    \includegraphics[width=0.88\textwidth]{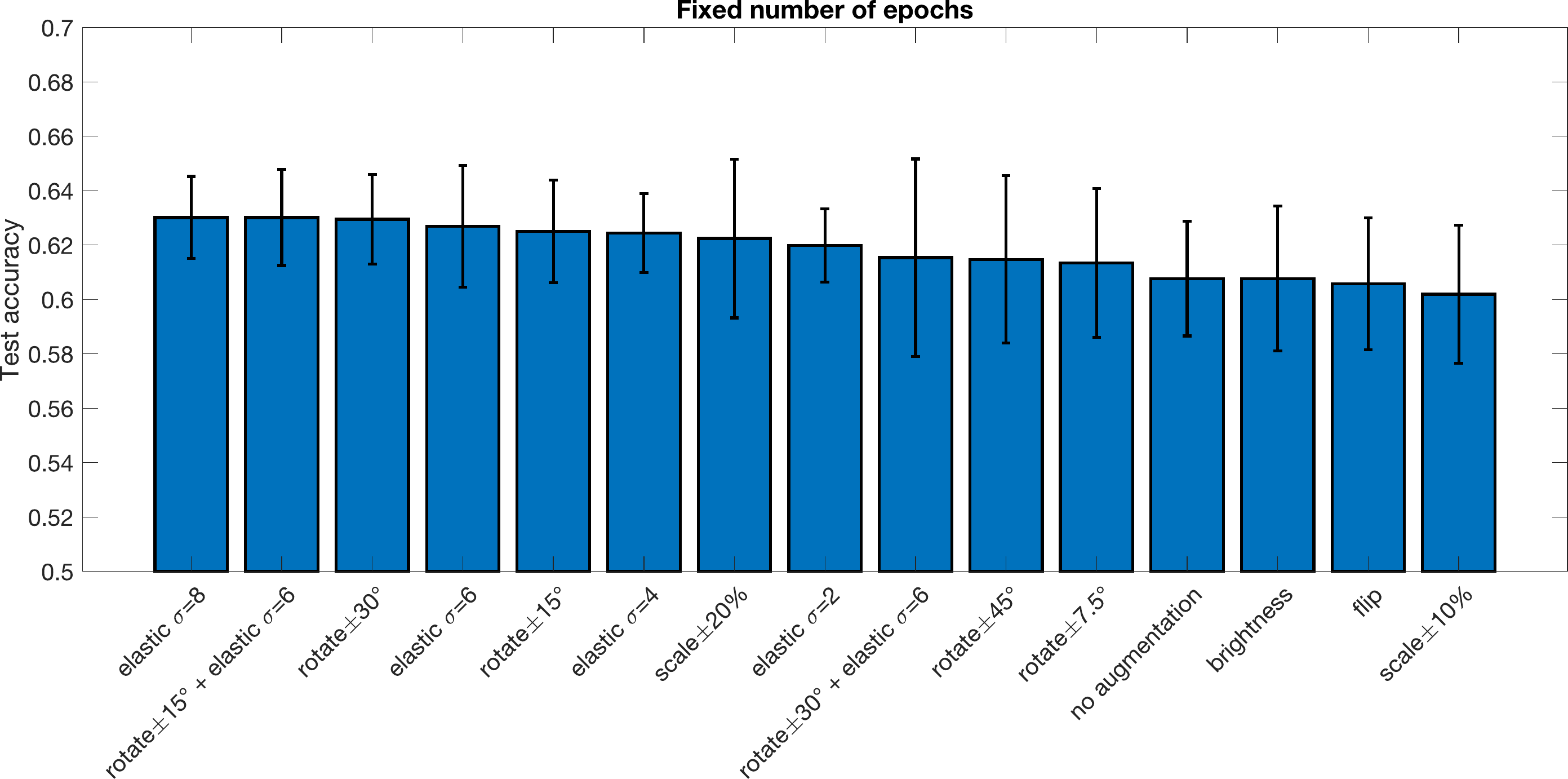}
    \caption{Test accuracy for classifying subjects as healthy or diseased for the ABIDE dataset, for different data augmentation approaches. The error bar represents the standard deviation over the 10 cross-validation folds. These results were obtained when using a fix number of epochs for each training. Compared to no augmentation, the best augmentation approach increases the test accuracy by 2.2 percentage units.}
    \label{fig:results2}
\end{figure*}

The results from all the different augmentation techniques, as well as baseline results obtained without augmentation, are presented in Figures~\ref{fig:results1} (early stopping) and~\ref{fig:results2} (fixed number of training epochs). In general the augmentation does not have a large effect on the test accuracy. For early stopping, random scaling seems to be the best single augmentation approach, but the mean improvement over 10 cross-validation folds is only about 0.5 percentage units. Small elastic deformations also have a small positive effect, while large deformations give the worst results. For a fixed number of training epochs, elastic deformations seem to work best, with an improvement of 2.2 percentage units.

\vspace{-0.25cm}
\section{Discussion}
\label{sec:discussion}

Compared to our previous work on augmentation for brain tumor segmentation~\cite{cirillo2021}, where several augmentation techniques were shown to significantly improve the segmentation accuracy, we only find minor improvements of the test accuracy in this study (even though the training accuracy is well above 90\%, indicating over fitting). Volume classification is in general a harder problem than volume segmentation, as each subject only represents a single training example, and may partly explain our results.

\clearpage

In this study brightness augmentation does not help at all, while it provided a major improvement for brain tumor segmentation for MR images collected at some 20 different sites~\cite{cirillo2021}. A possible explanation is that the data in this study are not raw MR images, since many preprocessing steps have been used to normalize the intensities to a certain range, and to calculate different derivatives. On the contrary, as the values in the derivative volumes have a certain meaning, brightness augmentation can impair the performance.

Since all the subjects have been registered to MNI space, we hypothesized that the results may be different if random transformations are applied to the test volumes, but test time augmentation did not change our findings (results not shown).

Our results are for a single preprocessing strategy (global signal regression and bandpass filtering), and the preprocessing choice can at least in theory affect how much the augmentation helps.

Our conclusion is that augmentation only provides minor improvements when training 3D CNNs for classification of ASD versus controls, but the results may be different for an easier task where the baseline test accuracy is for example 80\%. The results may also differ for other derivatives in ABIDE preprocessed, and when using several derivatives at the same time using a multi-channel 3D CNN.



%
%
%


\section{Compliance with ethical standards}
\label{sec:ethics}

This research study was conducted retrospectively using human subject data made available in open access by ABIDE preprocessed. Ethical approval was not required as confirmed by the ethics committee of Link\"{o}ping.
    
\section{Acknowledgments}
\label{sec:acknowledgments}

This work was supported by the Swedish research council grant 2017-04889, and by the ITEA / VINNOVA funded project Automation, Surgery Support and Intuitive 3D visualization to optimize workflow in IGT SysTems (ASSIST). Anders Eklund has previously received hardware from Nvidia, otherwise the authors have no conflicts of interest to declare.


\bibliographystyle{IEEEbib}
\bibliography{refs}

\end{document}